# A Radically New Theory of how the Brain Represents and Computes with Probabilities

Rod Rinkus, Neurithmic Systems, 468 Waltham St., Newton, MA 02465 USA
& Volen Center for Complex Systems, Brandeis University, 415 South St., Waltham, MA 02453 USA
Email: rod@neurithmicsystems.com     ORCID ID: 0000-0003-1725-910X

## Abstract

The brain is believed to implement probabilistic reasoning and to represent information via population, or distributed, coding. Most previous probabilistic population coding (PPC) theories share basic properties: 1) continuous-valued units; 2) fully/densely-distributed codes; 3) graded synapses; 4) rate coding; 5) units have innate unimodal tuning functions (TFs); 6) units are intrinsically noisy; and 7) noise/correlation is generally considered harmful. We present a radically different theory that assumes: 1) binary units; 2) only a small subset of units, i.e., a *sparse distributed representation* (SDR) (a.k.a. cell assembly, ensemble), comprises any individual code; 3) *functionally* binary synapses; 4) signaling formally requires only single (i.e., first) spikes; 5) units initially have completely flat TFs (all weights zero); 6) units are far less intrinsically noisy than traditionally thought; rather 7) noise is a resource generated/used to cause similar inputs to map to similar codes, controlling a tradeoff between storage capacity and embedding the input space statistics in the pattern of intersections over stored codes, epiphenomenally determining correlation patterns across neurons. The theory, Sparsey, was introduced 20+ years ago as a canonical cortical circuit/algorithm model achieving efficient spatiotemporal pattern learning/recognition, but was not elaborated as an alternative to PPC-type theories. Here, we show that: a) the active SDR simultaneously represents *both* the most similar/likely input *and* the entire (coarsely-ranked) similarity/likelihood distribution over *all* stored inputs (hypotheses); and b) given an input, Sparsey's code selection algorithm, which underlies both learning and inference, updates both the most likely hypothesis and the entire likelihood distribution (cf. belief update) with a number of steps that remains constant as the number of stored items increases.

## Acknowledgement

I'd like to thank the people who have encouraged me to pursue this theory over the years, including Dan Bullock, Jeff Hawkins, John Lisman, Josh Alspector, Tom McKenna, Andrew Browning, and Dan Hammerstrom. I'd also like to thank my colleagues at Neurithmic Systems, Greg Lesher, Jasmin Leveille, Oliver Layton, Harald Ruda, and Nick Nowak for their help developing these ideas. This work has been partially supported by DARPA contracts FA8650-13-C-7432 and N00173-09-C-2038, ONR contract N00014-12-C-0539, and NIH Training grant, 5 T32 NS07292.

# 1    Introduction

It is widely acknowledged that the brain must implement some form of probabilistic reasoning to deal with uncertainty in the world (Pouget, Beck et al. 2013).  However, exactly how the brain represents probabilities/likelihoods remains unknown (Ma and Jazayeri 2014, Pitkow and angelaki 2016).   It is also widely agreed that the brain represents information with some form of distributed—a.k.a. population, cell-assembly, ensemble—code [see (Barth and Poulet 2012) for relevant review].  Several population-based probabilistic coding theories (PPC) have been put forth in recent decades, including those in which the state of all neurons comprising the population, i.e., the *population code*, is viewed as representing: a) the single most likely/probable input value/feature (Georgopoulos, Kalaska et al. 1982); or b) the entire probability/likelihood distribution over features (Zemel, Dayan et al. 1998, Pouget, Dayan et al. 2000, Pouget, Dayan et al. 2003, Jazayeri and Movshon 2006, Ma, Beck et al. 2006, Boerlin and Denève 2011).  Despite their differences, these approaches share fundamental properties, a notable exception being the spike-based model of (Boerlin and Denève 2011).

1. Neural activation is continuous (graded).

2. *All* neurons in the coding field formally participate in the active code whether it represents a single hypothesis or a distribution over all hypotheses.  Such a representation is referred to as a *fully distributed r*epresentation.

3. Synapse strength is continuous (graded).

4. These approaches have generally been formulated in terms of rate-coding (Sanger 2003), which requires significant time, e.g., order tens of ms, for reliable decoding.

5. They assume *a priori* that tuning functions (TFs) of the neurons are unimodal, bell-shaped over any one dimension, and consequently do not explain how such TFs might develop through learning.

6. Individual neurons are assumed to be intrinsically noisy, e.g., firing with Poisson variability.

7. Noise and correlation are viewed primarily as problems that have to be dealt with, e.g., reducing noise correlation by averaging.

At a deeper level, it is clear that despite being framed as population models, they are really based on an underlying localist interpretation, specifically, that an individual neuron's firing rate can be taken as a perhaps noisy estimate of the probability that a single preferred feature (or preferred value of a feature) is present in its receptive field (Barlow 1972).  While these models entail some method of combining the outputs of individual neurons, e.g., averaging, each neuron is viewed as providing its own individual, i.e., *localist*, estimate of the input feature, i.e., each neuron possesses its own independent (typically bell-shaped) TF.  For example, this can be seen quite clearly in Fig. 1 of (Jazayeri and Movshon 2006) wherein the first layer cells (sensory neurons) are unimodal and therefore can be viewed as detectors of the value at their modes (preferred stimulus) and the *pooling* cells are also in 1-to-1 correspondence with directions.  This underlying localist interpretation is present in the other PPC models referenced above as well.

However, there are several compelling arguments against such localistically-rooted conceptions.  From an experimental standpoint, a growing body of research suggests that individual cell TFs are far more heterogeneous than classically conceived (Yen, Baker et al. 2007, Cox and DiCarlo 2008, Smith and Häusser 2010, Bonin, Histed et al. 2011, Mante, Sussillo et al. 2013, Nandy, Sharpee et al. 2013, Nandy, Mitchell et al. 2016), also described as having "mixed selectivity" (Fusi, Miller et al. 2016).  And, the greater the fidelity with which the heterogeneity of TFs is modeled, the less neuronal response variation that needs to be attributed to noise, leading some to question the appropriateness of the traditional concept of a single neuron I/O function as an invariant TF plus noise (Deneve and Chalk 2016).  From a formal standpoint, the limitation has long been pointed out that the maximum number of features/concepts, e.g., oriented edges, directions of movement, that can be stored in a localist coding field of $N$ units is $N$.  More importantly, as our own work has shown, the computational time efficiency with which features/concepts can be stored and retrieved, is far greater for memories in which items are represented with sparse distributed representations (SDRs) than for localist memories (Rinkus 1996, Rinkus 2010, Rinkus 2014).

The theory described herein, Sparsey, constitutes a radically new way of representing and computing with probabilities, diverging from most existing PPC theories in many fundamental ways, including:

1. The neurons comprising the coding field need only be *binary*.

2. Individual represented items / hypotheses are represented by fixed-size, sparsely chosen subsets of the population, i.e., SDRs (Rinkus 1996, Rinkus 2014).

3. Decoding (read-out) of the most likely hypothesis and of the entire distribution by downstream computations, requires only binary synapses.

4. Signaling can be communicated via a wave of contemporaneously arriving first-spikes (e.g., within a few ms window, likely organized by local gamma) from an afferent SDR code to a downstream computation (including recurrently to the source coding field) and is thus in principal, 1-2 orders of magnitude faster than rate coding. This means Sparsey is not formally a spiking model.

5. The initial weights of all afferent synapses to all neurons comprising the field are zero, i.e., the TFs are completely flat. Roughly unimodal TFs [as would be revealed by low-complexity probes, e.g., oriented bars spanning a cell's receptive field (RF)] emerge as a side-effect of the model's single/few-trial learning process of laying down SDRs in superposition *provided that the process of choosing those SDRs preserves similarity* (which Sparsey's learning algorithm does).

6. Neurons are far less inherently noisy than has been classically supposed. Most observed noise has likely been due to experimental limitations, i.e., the inability to truly closely control input conditions. At an algorithmic, or information-bearing level, individual neurons are simply on or off on any given computational cycle. We make no use of precise spike times, but assume that some meta-circuitry organizes (temporally collects) transmission of *en masse* synaptic signals from an SDR coding field within some small window of arrival at a downstream decoding field. Our working assumption is that this is the purpose of the local gamma cycle (Fries 2009, Buzsáki 2010, Igarashi, Lu et al. 2014, Watrous, Fell et al. 2015).

7. While neurons are inherently less noisy than has been assumed, the canonical, mesoscale (i.e., the cell assembly scale) circuit does explicitly use noise as a resource. That is, noise (presumably mediated by neuromodulators, e.g., NE, ACh) is explicitly generated and injected into the code selection process to achieve a specific coding goal. That goal is to make the SDR code that is activated in response to an input be increasingly random as a function of novelty, i.e., to minimize the intersection of the activated code with all previously stored codes. More generally, the goal is that the overall set of codes stored in a coding field has the property that similar inputs map to similar codes ("SISC"), where the similarity measure for SDR codes is size of intersection. This has straightforward implications on correlation, described below.

SDR admits a completely different concept for representing and computing with probabilities than is possible under either localist or fully distributed representations. Instead of representing an input/feature, X, by the state of a *single* neuron as in localism, or by a vector of real values over *all* neurons comprising the coding field, as in fully distributed coding (e.g., as in a hidden level of a multi-layer perceptron model), in SDR, X is represented by a *subset* of neurons, specifically, a small subset of the whole coding field, which we refer to as X's code. All neurons participating in the code are fully active (in a binary sense) and the rest of the coding field's neurons are completely off, which is closely consistent with the combinatorial coding framework described and analyzed in (Osborne, Palmer et al. 2008). Thus:

A. Gradations in the certainty, or probability, that X is present in the receptive field (RF) of the SDR coding field can be represented by different *fractions* of X's code being active (see Fig. 2).

B. Communication of X's probability to target (downstream, or subsequent) computations—i.e., *decoding*—can be achieved by neurons participating in such target computations simply *summing* their binary synaptic inputs from the (source) field in which X is active.

Thus, representing and using graded values, e.g., probabilities, requires only binary neurons and binary synapses. There is no need for explicit *localist* representation of graded values, anywhere in the system/computation. More specifically, there is no need to represent probabilities/likelihoods localistically in the form of firing rates (nor in the form of exact spike times within some window) and there is no need to represent conditional probabilities, or strengths of association, via continuous/graded weights. Moreover, although neurons presumably communicate primarily via spikes, conceiving the signal sent from an active SDR code as a vector of contemporaneously arriving first spikes (to a target field) essentially removes the need to cast our treatment in terms of a "spiking model" with its assumed (typically Poisson) intrinsic noisiness. Rather, our approach has more the character of an algorithm operating according to a discrete clock (whose basic cycle, we suppose is a gamma cycle). Thus, our approach does not assume intrinsic noise 'that has to be dealt with', but rather, as we'll describe, injects a state-dependent amount of noise by manipulating neurons' transfer functions, specifically, their nonlinearities, to achieve certain coding goals, as discussed below. This underscores a strongly distinguishing property of Sparsey. The nonlinearity of the principal cells is not static as is true of most models, but highly dynamic: specifically, the nonlinearities of all principal cells comprising a coding field are modulated *en masse*, in correlated fashion, and on a fast time scale, e.g., ~10 ms, as a function of a global (to the coding field) measure of the familiarity (inverse novelty) of the total input to the coding field.

The unique property of SDR that the similarity of represented items/features can be represented by the intersection size of their codes, i.e., SISC, has been pointed out in the past (Palm, Schwenker et al. 1995, Rinkus 1996, Rachkovskij and Kussul 2001, Kanerva 2009). However, the possibility of:

a) interpreting the fraction of a feature's SDR code that is active as the probability/likelihood of that feature (hypothesis), and

b) interpreting the code active in an SDR coding field, whether it be a reactivation of a previously stored code or a novel code (which nevertheless, may generally overlap with many previously stored codes), as a *distribution over all stored codes (and thus, over all represented inputs/features) stored in the field* was introduced in (Rinkus 2012). Similarity preservation, involving a continuous measure, e.g., Euclidean distance, has also been established for fully distributed coding models, e.g., (Bogacz 2007). However, to our knowledge, such codes have not been interpreted as *simultaneously* representing *both* the most likely hypothesis *and* the entire distribution. In any case, as noted above, in such models, decoding by downstream computations requires either graded synapses or rate-coding.

Above, we refer to the "RF of the SDR coding field" rather than to the RF of an individual neuron. This is because we require that, to first approximation, all neurons in the SDR coding field have the same RF in order to assert that the codes in that field represent features present in the RF. This constraint can be relaxed to some extent, but it facilitates initial exposition and analysis. In any case, this constraint actually serves to further underscore the massive gulf between Sparsey and previous PPC theories for which this constraint is not needed, even approximately, due to their underlying localist conception. Sparsey evinces a new paradigm of neuroscience in which the ensemble (SDR, cell assembly), not the single neuron, is considered the fundamental functional unit, i.e. a move away from the 'Neuron Doctrine' as for example, advocated in (Yuste 2015, Fusi, Miller et al. 2016, Schneidman 2016). We expect that with rapidly advancing methods [recent reviews: (Hamel, Grewe et al. 2015, Jercog, Rogerson et al. 2016)] allowing observation of the activities of all neurons in substantial (e.g., order hypercolumn/barrel-sized) volumes and with progressively greater temporal precision, issues such as our connectivity assumption as well as the dynamics implied in our core algorithm will be addressable.

In the results section, we present two simulation-backed examples showing the mechanistic details of how Sparsey's core algorithm, the *Code Selection Algorithm* (CSA) (Rinkus 1996, Rinkus 2004, Rinkus and Lisman 2005, Rinkus 2008, Rinkus 2010, Rinkus 2013, Rinkus 2014), activates not only the code of the best-matching, or most likely, input, but also activates the *entire* similarity/likelihood distribution, with coarsely-ranked fidelity, over *all* stored inputs. Moreover, the CSA accomplishes this with a number of steps that remains constant as the number of stored inputs increases, a property that we call *fixed* time complexity. The first example shows this for the case of purely spatial inputs. The second example shows that when the coding field is recurrently connected, the entire distribution is updated from time T to T+1 [which can be viewed as *belief update* (Pearl 1988)] in a way that is consistent with the statistics of the domain, again, in *fixed* time. Moreover, as explained previously (Rinkus 2010, Rinkus 2014), the CSA also stores, or learns, new inputs, spatial or spatiotemporal, in fixed time. In fact, Sparsey/CSA can be viewed as an adaptive hashing method which learns a *locality-sensitive*, i.e., similarity-preserving, hash function from the data (which can be spatial or spatiotemporal). While other neurally inspired hashing models have fixed-time best-match *retrieval* (Salakhutdinov and Hinton 2007, Salakhutdinov and Hinton 2009, Grauman and Fergus 2013), they *do not* also have fixed-time *learning*. In fact, none of the currently state-of-art hashing models described in a recent review (Wang, Liu et al. 2016) possess both fixed-time learning and fixed-time best-match retrieval. Although time complexity considerations like these have generally not been discussed in the probabilistic population coding literature, they are essential for evaluating the overall plausibility of models of biological cognition, for while it is uncontentious that the brain computes probabilistically, we also need to explain the extreme speed with which these computations, which are over potentially quite large hypothesis spaces, occur.

A crucial reason why Sparsey achieves fixed time performance for both learning and best-match retrieval is its unique and simple method for computing the *global familiarity*, $G$, of an input and using it to control the selection of a code for that input. By 'global', we mean that $G$ is a function of all cells comprising the coding field, in contrast to *local familiarity*, $V_i$, which is cell $i$'s local (depending only on its synaptic inputs) measure of its match to the input. Crucially, computing $G$ *does not require* explicitly comparing the new input to every stored input (nor to a log number of the stored inputs as is the case for tree-based methods). Rather, the $G$ computation, whose time complexity is dominated by a single pass over the *fixed* number of afferent weights to the field, implicitly performs these comparisons simultaneously, in an algorithmically parallel fashion. 'Algorithmic parallelism' means that single atomic operations affect multiple represented (stored) items. Thus, operationally, 'algorithmic parallelism' is very close if not identical to 'distributed representation': you cannot have one without the other. We emphasize that algorithmic parallelism and machine parallelism are orthogonal resources and are fully compatible.

A simplified version of the CSA, sufficient for this paper's examples, is given in Table 1, but we briefly summarize it here. CSA Step 1 computes the input sums for all $Q{\times}K$ cells comprising the coding field. Specifically, for each cell, a separate sum is computed for each of its major afferent synaptic projections, e.g., bottom-up (U), horizontal (H), and top-down (D) projections, the latter two of which provide recurrence to the coding field. This is the step requiring the aforementioned single pass over the field's afferent weights. In Step 2, these sums are normalized and in Step 3, the normalized sums are (optionally nonlinearly transformed and) multiplied, yielding the $V$ values. In Steps 4 and 5, $G$ is computed as the average max $V$ across the $Q$ CMs. In the remaining CSA steps, $G$ is used, in each CM, to nonlinearly transform the $V$ distribution over the $K$ cells into a final $\rho$ distribution from which a winner is picked. $G$'s influence on the distributions can be summarized as follows.

a) When high global familiarity is detected ($G{\approx}1$), those distributions are exaggerated to bias the choice in favor of cells that have high input summations, and thus, high *local* familiarities, $V_i$, which acts to increase correlation.

b) When low global familiarity is detected ($G{\approx}0$), those distributions are flattened so as to reduce bias due to local familiarity, which acts to increase the expected Hamming distance between the selected code and previously stored codes, i.e., to decrease correlation.

Since the $V$ values represent *signal*, exaggerating the $V$ distribution in a CM increases signal whereas flattening it increases noise. The above behavior (and its smooth interpolation over the range, $G=1$ to $G=0$) is the means by which Sparsey achieves SISC. And, it is the enforcement (statistically) of SISC during learning, which ultimately makes possible the immediate (fixed time complexity) retrieval of the best-matching (or most likely, most relevant) hypothesis and the simultaneous fixed-time update of *all* stored hypotheses (i.e., via algorithmic parallelism, not machine parallelism) with each new input.

## 1.1    Novel explanation of noise and correlation

In recent years, there has been much discussion about the nature, causes, and uses, of correlations and noise in cortical activity; see (Cohen and Kohn 2011, Kohn, Coen-Cagli et al. 2016, Schneidman 2016) for reviews. Most investigations of neural correlation and noise, especially in the context of PPC theories, assume *a priori*: a) fundamentally noisy neurons, e.g., Poisson spiking; and b) tuning functions (TFs) of some general form, e.g., unimodal, bell-shaped, and then describe how noise/correlation affects the coding accuracy of populations with such TFs (Abbott and Dayan 1999, Moreno-Bote, Beck et al. 2014, Franke, Fiscella et al. 2016, Rosenbaum, Smith et al. 2017). Specifically, these treatments measure correlation in terms of either mean spiking rates ("signal correlation") or spikes themselves ("noise correlations"). However, as noted above, our theory makes neither assumption. Rather, in our theory, noise (randomness) is *actively injected*—implemented via the *G*-dependent modulation of the neuronal transfer function—during learning to achieve the goals described above. Thus, the pattern of correlations amongst units (neurons) simply emerges as a side-effect of cells being selected to participate in codes. The overarching goal of the learning process is simply to enforce SISC. However, enforcing SISC in the context of an SDR coding field realizes a balance between:

a)  maximizing the storage capacity of the coding field, and

b)  embedding the similarity structure of the input space in the set of stored codes, which in turn enables fixed-time best-match retrieval.

Interestingly, in exploring the implications of shifting focus from information theory to coding theory in terms of influence upon theoretical neuroscience, (Curto, Itskov et al. 2013) have pointed to this same tradeoff, though their treatment uses error rate (coding accuracy) instead of storage capacity. We point out that understanding how neural correlation ultimately affects things like storage capacity is considered largely unknown and an active area of research (Latham 2017). Our approach implies a straightforward answer. Minimizing correlation, i.e., maximizing average Hamming distance over the set of codes stored in an SDR coding field, maximizes storage capacity. Increases of any correlations of pairs, triples, or subsets of any order, of the coding field's cells decreases capacity.

## 2   Results

### 2.1   A single SDR code represents an entire similarity/probability/likelihood distribution: Conceptual introduction

Fig. 1a shows Sparsey's particular SDR format. The coding field consists of *Q* winner-take-all (WTA) competitive modules (CMs), each consisting of *K* binary neurons. Here, *Q*=7 and *K*=7. Thus, all codes have exactly *Q* active neurons and there are $K^Q$ possible codes. We assume that the input field of binary neurons (hexagons), e.g., representing an 8x8 pixel patch of visual field, is *completely connected* to the coding field (blue lines represent weights), i.e., the aforementioned "RF of SDR coding field". All weights are initially zero. Fig. 1b shows a particular input A, which has been associated with a particular code, *ϕ*(A); here, blue lines indicate the bundle [cf. "Synapsemble", (Buzsáki 2010)] of weights that would be increased from 0 to 1 to store this association (memory trace).

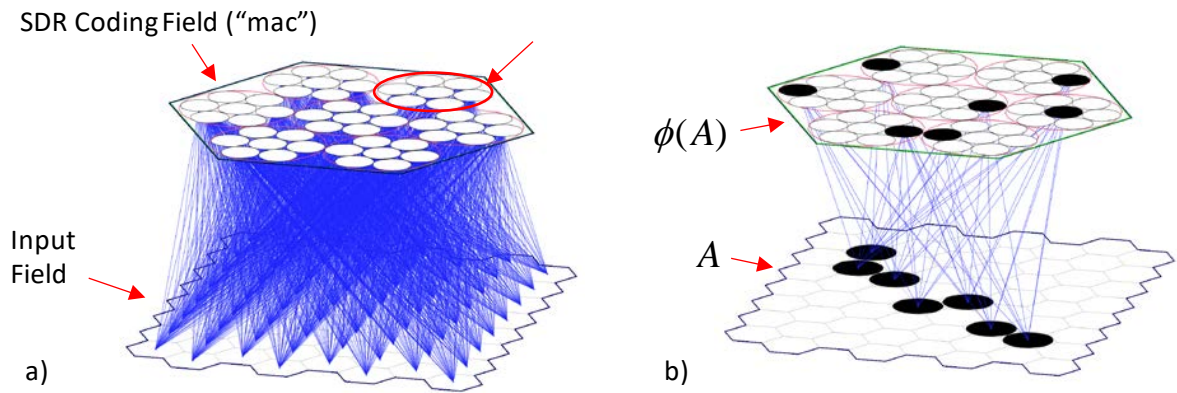

**Fig. 1** a) The sparse distributed representation (SDR) coding field (macrocolumn, "mac") consists of *Q*=7 WTA competitive modules. An input field of binary neurons is completely connected to the coding field. b) A particular input A, its code, *ϕ*(A), consisting of *Q* active (black) neurons, and the bundle of weights that would be increased to form the association are shown

Fig. 2 shows how the strength of presence, i.e., (posterior) probability, of a feature in the coding field's RF is represented in our SDR-based theory. The figure shows five hypothetical inputs, A-E, which have been learned, i.e., associated with codes, *ϕ*(A) - *ϕ*(E). We hand-chose these particular codes to be consistent with the principle that *similar inputs* should map to *similar codes* (SISC). That is, inputs B to E have progressively smaller overlaps with A and therefore codes *ϕ*(B) to *ϕ*(E) have progressively smaller intersections with *ϕ*(A). The CSA has been shown to statistically enforce SISC for both the spatial and spatiotemporal (sequential) input domains (Rinkus 1996, Rinkus 2008, Rinkus 2010, Rinkus 2013, Rinkus 2014).

For input spaces for which it is plausible to assume that input similarity correlates with probability/likelihood, the single active code can therefore also be viewed as a probability/likelihood distribution over all stored codes. This is shown in the lower part of the figure. The leftmost panel at the bottom of Fig. 2 shows that when *ϕ*(A) is 100% active, the other codes are partially active in proportions that reflect the similarities of their corresponding inputs to A, and thus the probabilities/likelihoods of the inputs they represent. The remaining four panels show input similarity (probability/likelihood) approximately correlating with code overlap.

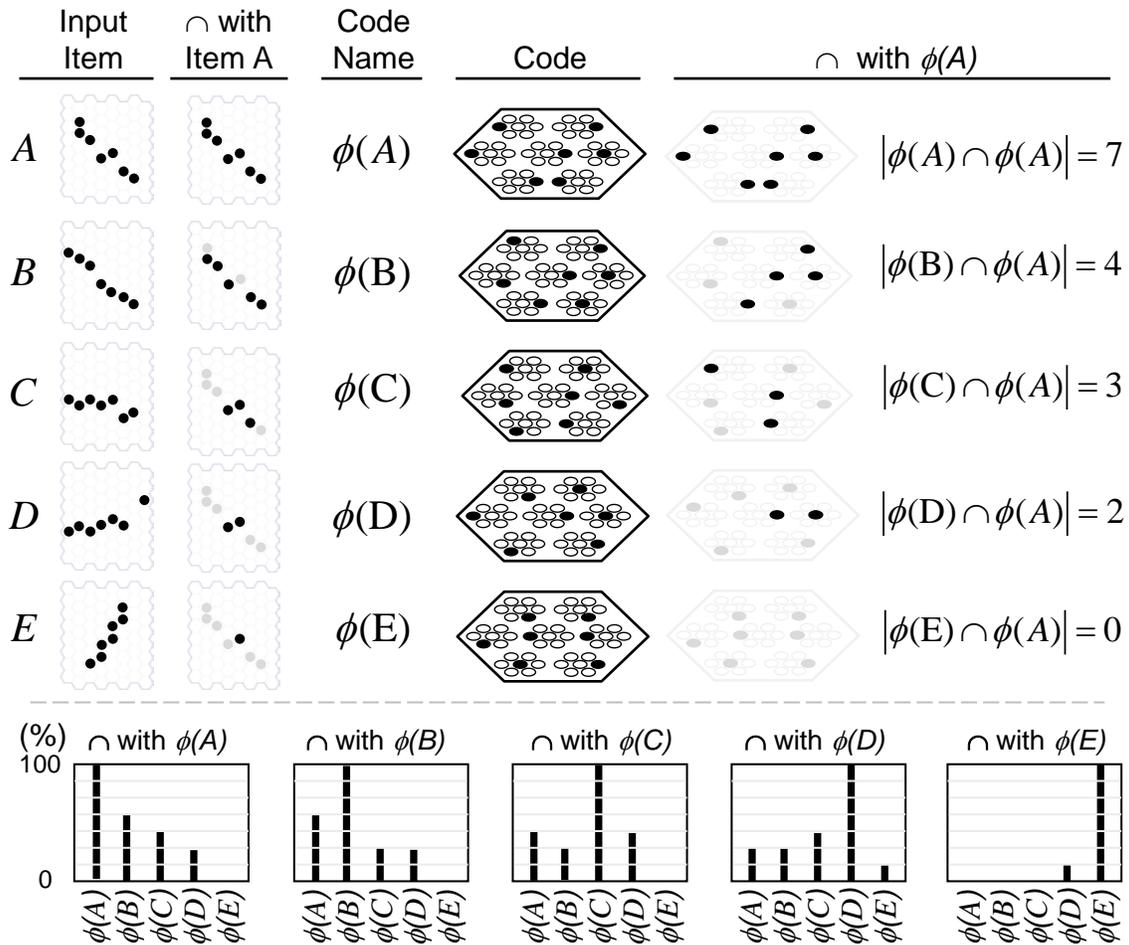

**Fig. 2** Illustration of how the probability/likelihood of a feature can be represented by the fraction of its code that is active. When $\phi$(A) is fully active, the hypothesis that feature A is present can be considered maximally probable. Because the similarities of the other features to the most probable feature, A, correlate with their codes' overlaps with $\phi$(A), the probabilities/likelihoods of those features are represented by the fraction of their codes that are active. In the intersection ("∩") columns, black indicates units intersecting with either the input pattern A or its code, $\phi$(A); gray indicates non-intersecting units

The example of Fig. 2 was constructed to illustrate the desired property that the similarities, and thus, likelihoods, of all stored hypotheses are simultaneously physically active whenever any single hypothesis is fully active. The next section demonstrates that the CSA achieves this property for purely spatial inputs and the following section, for the spatiotemporal case. Table 1 presents a simplified version of the code selection algorithm (CSA) with the minimal steps needed for these demonstrations. Specifically, the simplifications, relative to (Rinkus 2014) are: a) we use a model with only one internal level, thus there are no D signals; b) all spatial inputs have exactly 12 active pixels, thus the U normalizer can be constant, $\pi_U = 12$; and c) the internal level consists of a single coding field (mac), thus the H normalizer can be constant, $\pi_H = Q - 1$.

**Table 1** Simplified Code Selection Algorithm (CSA)

| | Equation | Short Description |
|---|---|---|
| 1 | $u(i) = \sum_{j \in RF_U} x(j) w(j,i)$ <br> $h(i) = \sum_{j \in RF_H} x(j,t-1) w(j,i)$ | Compute raw U (and H, if applicable) input sums. |
| 2 | $U(i) = u(i) \big/ \pi_U \, w_{max}$ <br> $H(i) = h(i) \big/ \pi_H \, w_{max}$ | Compute normalized U (and H, if applicable) input sums. In this paper's simulations, $\pi_U = 12$ and $\pi_H = Q - 1$. |
| 3 | $V(i) = \begin{cases} H(i)^{\lambda_H} U(i)^{\lambda_U} & t \geq 1 \\ U(i)^{\lambda_U} & t = 0 \end{cases}$ | Compute local evidential support for each cell. In this paper, $\lambda_H = \lambda_U = 1$, unless otherwise stated. |
| 4 | $\hat{V}_j = \max_{i \in C_j} \{V(i)\}$ | Find the max V, $\hat{V}_j$, in each CM, $C_j$ |
| 5 | $G = \sum_{q=1}^{Q} \hat{V}_k \big/ Q$ | Compute G as average $\hat{V}$ value over the Q CMs |
| 6 | $\eta = 1 + \left( \left[ \dfrac{G - G^-}{1 - G^-} \right]^+ \right)^{\gamma} \times \chi \times K$ | Determine expansivity ($\eta$) of V-to-$\mu$ sigmoid function. In this paper, $\gamma=2$, $\chi=100$, $G^- = 0.1$ |
| 7 | $\sigma_1 = \dfrac{\left( (\eta - 1) / 0.001 \right)^{1/\sigma_4} - 1}{e^{\sigma_2 \sigma_3}}$ | Sets $\sigma_1$ so that the overall sigmoid shape is preserved over full $\eta$ range. <br> $\sigma_2 = 7.0$, $\sigma_3 = 0.4$, $\sigma_4 = 9.5$ |
| 8 | $\mu(i) = \dfrac{(\eta - 1)}{\left( 1 + \sigma_1 e^{-\sigma_2 (V(i) - \sigma_3)} \right)^{\sigma_4}} + 1$ | To each cell, apply sigmoid function, which collapses to constant fn, $\mu(i) = 1$, when $G \leq G^-$) |
| 9 | $\rho(i) = \dfrac{\mu(i)}{\sum_{k \in CM} \mu(k)}$ | In each CM, normalize relative ($\mu$) to final ($\rho$) probabilities of winning |
| 10 | Select a final winner in each CM according to the $\rho$ distribution in that CM, i.e., soft max. | |

## 2.2 Single SDR code represents entire probability/likelihood distribution: Spatial case

Fig. 3a shows six inputs, $I_1$ to $I_6$ (which are disjoint for simplicity of exposition) that have been previously stored in the model instance depicted in Fig. 3b. The model has a 12x12 binary pixel input level that is completely connected to all cells comprising the mac. The mac consists of $Q$=24 WTA CMs, each having $K$=8 binary cells. The second row of Fig. 3a shows a novel test stimulus, $I_7$, and its varying overlaps (yellow pixels) with $I_1$ to $I_6$. Given that all inputs are constrained to have exactly 12 active pixels, we can measure spatial similarity, $sim(I_x, I_y)$, simply as size of intersection divided by 12 (shown as decimals under inputs):

$$sim(I_x, I_y) = \left| I_x \cap I_y \right| / 12$$

Fig. 3b shows the code, $\phi(I_7)$, activated in response to $I_7$, which by construction is most similar to $I_1$. Black coding cells are cells that also won for $I_1$, red indicates active cells that did not win for $I_1$, and green indicates inactive cells that did win for $I_1$. The red and green cells in a given CM can be viewed as a substitution error. The intention of the red color for coding cells is that if this is a retrieval trial in which the model is being asked to return the closest matching stored input, $I_1$, then the red cells can be considered errors. Note however that these are sub-symbolic scale errors, not errors at the scale of whole inputs (hypotheses, symbols), as whole inputs are *collectively* represented by the entire SDR code. In this example appropriate threshold settings in downstream/decoding units, would allow the model as a whole return the correct answer given that 18 out of 24 cells of $I_1$'s code, $\phi(I_1)$, are activated, similar to thresholding schemes in other associative memory models (Marr 1969, Willshaw, Buneman et al. 1969). Note however that if this was a learning trial, then the red cells would not be considered errors: this would simply be a new code, $\phi(I_7)$, being assigned to represent a novel input, $I_7$, and in a way that respects similarity in the input space.

Fig. 3d shows the first key message of the figure. The active fractions of the codes, $\phi(I_1)$ to $\phi(I_6)$, representing the six stored inputs, $I_1$ to $I_6$, are highly rank-correlated with the pixel-wise similarities of these inputs to $I_7$. Thus, the blue bar in Fig. 3d represents the fact that the code, $\phi(I_1)$, for the best matching stored input, $I_1$, has the highest active code fraction, 75% (18 out 24, the black cells in Fig. 3b) cells of $\phi(I_1)$ are active in $\phi(I_7)$. The cyan bar for the next closest matching stored input, $I_2$, indicates that 12 out of 24 of the cells of $\phi(I_2)$ (code note shown) are active in $\phi(I_7)$. In general, many of these 12 may be common to the 18 cells in $\left\{ \phi(I_7) \cap \phi(I_1) \right\}$. And so on for the other stored hypotheses. (The actual codes, $\phi(I_1)$ to $\phi(I_6)$, are not shown; only the intersection sizes with $\phi(I_7)$ matter and those are indicated along right margin of chart in Fig. 3d.) We note that even the code for $I_6$ which has zero intersection with $I_7$ has two cells in common with $\phi(I_1)$. In general, the expected code intersection for the zero input intersection condition is not zero, but chance, since in that case, the winners are chosen from the uniform distribution in each CM: thus, the expected intersection for that case is just $Q/K$.

As noted earlier, we assume that the similarity of a stored input $I_X$ to the current input can be taken as a measure of $I_X$'s probability/likelihood. And, since all codes are of size $Q$, we can divide code intersection size by $Q$, yielding a measure normalized to [0,1]:

$$L(I_1) = \left| \phi(I_7) \cap \phi(I_1) \right| / Q$$

We also assume that $I_1$ to $I_6$ each occurred exactly once during training and thus, that the prior over hypotheses is flat. In this case the posterior and likelihood are proportional to each other, thus, the likelihoods in Fig. 3d can also be viewed as unnormalized posterior probabilities of the hypotheses corresponding to the six stored codes.

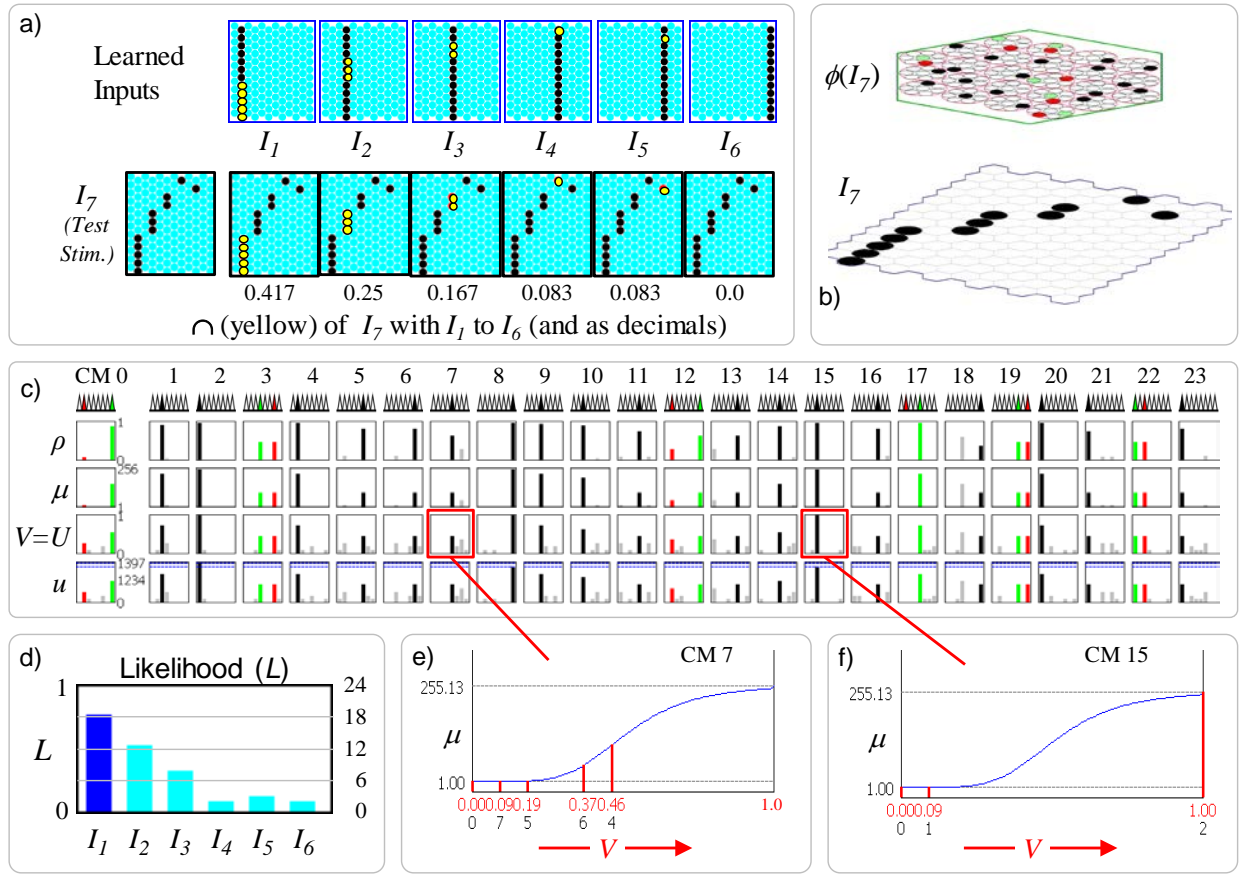

**Fig. 3** In response to an input, the codes for learned (stored) inputs, i.e., hypotheses, are activated with strength that is correlated with the similarity (pixel overlap) of the current input and the learned input. Test input $I_7$ is most similar to learned input $I_1$, shown by the intersections (yellow pixels) in panel a. Thus, the code with the largest fraction of active cells is $\phi(I_1)$ (18/24=75%) (blue bar in panel d). The other codes are active in rough proportion to the similarities of $I_7$ and their associated inputs (cyan bars). (c) Raw ($u$) and normalized ($U$) input summations to all cells in all CMs. $V$ values, which equal the $U$ values in this purely spatial case, are transformed to un-normalized win probabilities ($\mu$) in each CM via sigmoid transform whose properties, e.g., max value of 255.13, depend on $G$ and other parameters. $\mu$ values are normalized to true probabilities ($\rho$) and one winner is chosen in each CM (indicated in row of triangles: black: winner for $I_7$ that also won for $I_1$; red: winner for $I_7$ that did not win $I_1$: green: winner for $I_1$ that did not win for $I_7$. (e, f) Details for CMs, 7 and 15. Values in second row of $V$ axis are indexes of cells having the $V$ values above them. Some CMs have a single cell with much higher $V$ and ultimately $\rho$ value than the rest (e.g., CM 15), some others have two cells that are tied for the max (e.g., CMs 3, 19, 22)

We acknowledge that the likelihoods in Fig. 3d may seem high. After all, $I_7$ has less than half its pixels in common with $I_1$, etc. Given these particular input patters, is it really reasonable to consider $I_1$ to have such high likelihood? Bear in mind that our example assumes that the only experience this model has of the world are single instances of the six inputs shown. We assume no prior knowledge of any underlying statistical structure generating the inputs. Thus, it is really only the *relative* values that matter and we could pick other parameters, notably in CSA Steps 6-8, that would result in a much less expansive sigmoid nonlinearity, which would result in lower expected intersections of $\phi(I_7)$ with the learned codes, and thus lower likelihoods. The main point is simply that the expected code intersections correlate with input similarity, and thus, likelihood.

Fig. 3c shows the second key message: the likelihood-correlated pattern of activation levels of the codes (hypotheses) apparent in Fig. 3d is achieved via independent soft max choices in each of the $Q$ CMs. Fig. 3c shows, for all 196 mac cells, the traces of the relevant variables used to determine $\phi(I_7)$. The raw input summation from active pixels is indicated by $u$. In this paper, all weights are effectively binary, though "1" is represented with 127 and "0" with 0. Hence, the maximum $u$ value possible in any cell when $I_7$ is presented is 12x127=1524. The model in this example also assumes that all inputs will have exactly 12 active pixels (this can be relaxed but is assumed for simplicity). $U$ is the normalized $u$ value, as in Eq. 2, where $\pi_U = 12$. We assume $\lambda_U = 1$ here, thus by Eq. 3, $V_i = U_i$. A cell's $V$ value represents the total *local evidence* that it should be activated. However, rather than simply picking the max $V$ cell in each CM as winner (i.e., hard max), which would amount to executing only steps 1-4 of the CSA, the remaining CSA steps, 5-10, are executed, in which the $V$ distributions are transformed as described earlier and winners are chosen via soft max in each CM [final winner choices, chosen from the $\rho$ distributions are shown in the row of triangles just below CM indexes]. Thus, an extremely cheap-to-compute (CSA Step 5) *global* function of the whole mac, $G$, is used to influence the *local* decision process in each CM. We repeat for emphasis that no part of the CSA explicitly operates on, i.e., iterates over, stored hypotheses; indeed, there are no explicit (localist) representations of stored hypotheses on which to operate.

Fig. 4 shows that different inputs yield different likelihood distributions that correlate approximately with similarity. Input $I_8$ (Fig. 4a) has highest intersection with $I_2$ and a different pattern of intersections with the other learned inputs as well (refer to Fig. 3a). Fig. 4c shows that the codes of the stored inputs become active in approximate proportion to their similarities with $I_8$, i.e., their likelihoods are simultaneously physically represented by the fractions of their codes which are active. The $G$ value in this case, 0.65, yields, via CSA steps 6-8, the $V$-to-$\mu$ transform shown in Fig. 4b, which is applied in all CMs. Its range is [1,300] and given the particular $V$ distributions shown in Fig. 4d, the cell with the max $V$ in each CM ends up being greatly favored over other lower-$V$ cells. The red box shows the $V$ distribution for CM 9. The second row of the abscissa in Fig. 4b gives the within-CM indexes of the cells having the corresponding (red) values immediately above (shown for only three cells). Thus, cell 3 has $V$=0.74 which maps to approximately $\mu \approx 250$ whereas its closest competitors, cells 4 and 6 (gray bars in red box) have $V$=0.19 which maps to $\mu = 1$. Similar statistical conditions exist in most of the other CMs. However, in three of them, CMs 0, 10, and 14, there are two cells tied for max $V$. In two, CMs 10 and 14, the cell that is not contained in $I_2$'s code, $\phi(I_2)$, wins (red triangle and bars), and in CM 0, the cell that is in $\phi(I_2)$ does win (black triangle and bars). Overall, presentation of $I_8$ activates a code $\phi(I_8)$ that has 21 out of 24 cells in common with $\phi(I_2)$ manifesting the high likelihood estimate for $I_2$.

To finish our demonstration for the spatial input case, Fig. 4e shows presentation of another input, $I_9$, having half its pixels in common with $I_3$ and the other half with $I_6$. Fig. 4g shows that the codes for $I_3$ and $I_6$ have both become approximately equally (with some statistical variance) active and are both more active than any of the other codes. Thus, the model is representing that these two hypotheses are the most likely and approximately equally likely. The exact bar heights fluctuate somewhat across trials, e.g., sometimes $I_3$ has higher likelihood than $I_6$, but the general shape of the distribution is preserved. The remaining hypotheses' likelihoods also approximately correlate with their pixelwise intersections with $I_9$. The qualitative difference between presenting $I_8$ and $I_9$ is readily seen by comparing the $V$ rows of Fig. 4d and 4h and seeing that for the latter, a tied max $V$ condition exists in almost all the CMs, reflecting the equal similarity of $I_9$ with $I_3$ and $I_6$. In approximately half of these CMs the cell that wins intersects with $\phi(I_3)$ and in the other half, the winner intersects with $\phi(I_6)$. In Fig. 4h, the three CMs in which there is a single black bar, CMs 1, 7, and 12, indicates that the codes, $\phi(I_3)$ and $\phi(I_6)$, intersect in these three CMs.

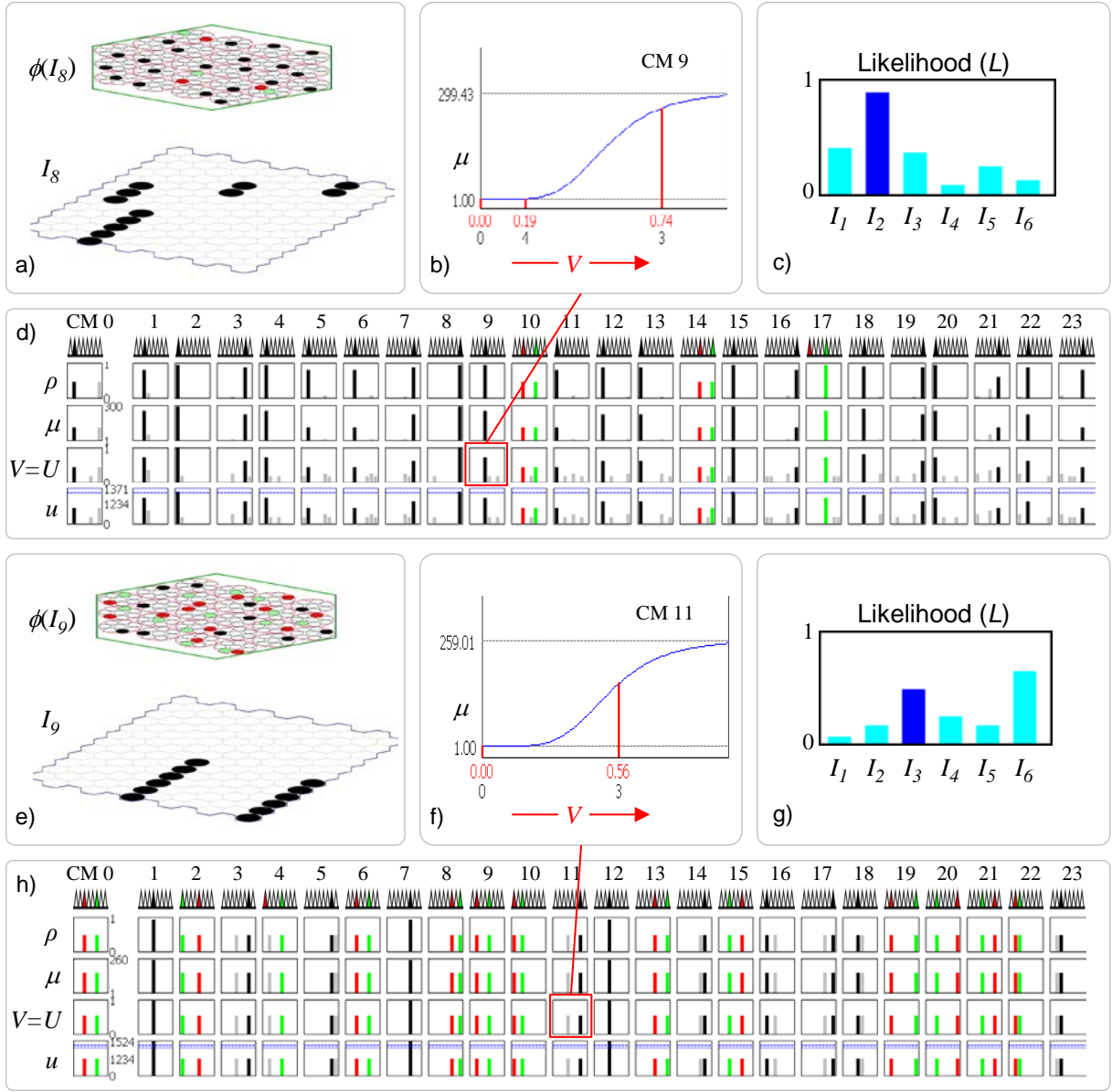

**Fig. 4** Details of presenting two further novel inputs, $I_8$ (panels a-d) and $I_9$ (panels e-h). In both cases, the resulting likelihood distributions correlate closely with the input overlap patterns. Panels b and f show details of one example CM (indicated by red boxes in panels d and h) for each input

### 2.3    Single SDR code represents entire prob. distribution: Spatiotemporal (sequential) case

Our goal in this section is to demonstrate moment-to-moment (frame-by-frame) updating of the similarity/likelihood distribution over stored inputs, which in this case are particular spatiotemporal moments, in approximate agreement with spatiotemporal similarity structure of the experienced inputs. Fig. 5 (top row) shows the training set consisting of two 4-item sequences, S1=[ABCD] and S2=[EFGH], where the items are the 12x12 pixel patterns shown. Fig. 5 (bottom row) shows two novel test sequences, S3 and S4, constructed from the same or slightly perturbed versions of the frames comprising the training sequences. We will present the details of testing on S3 and S4, but begin by showing, as a baseline, the details of testing on a training sequence, S1, in Figures 6 and 7.

The training sequences were handcrafted to show naturalistic edge motion patterns [assuming simple preprocessing (edge-filtering, binarization, and skeletonization)] on a 12x12 aperture of the visual field and such that the degree of pixel overlap amongst the frames is low. The test sequences were constructed so that their first and second halves would unambiguously be spatiotemporally most similar, in terms of the raw measure, pixel overlap, to the first and second halves of the training sequences. Thus, the S3 subsequence [A'B] is clearly most similar to S1 subsequence [AB], S3 subsequence [GH] is clearly most similar (in fact identical) to S2 subsequence [GH], etc. The overall goal of the demonstrations of S3 and S4 is to show that the likelihood distribution shifts to reflect, approximately, the spatiotemporal similarities of the stored hypotheses as we switch at mid-sequence between slightly noisy/perturbed versions of the two learned sequences.

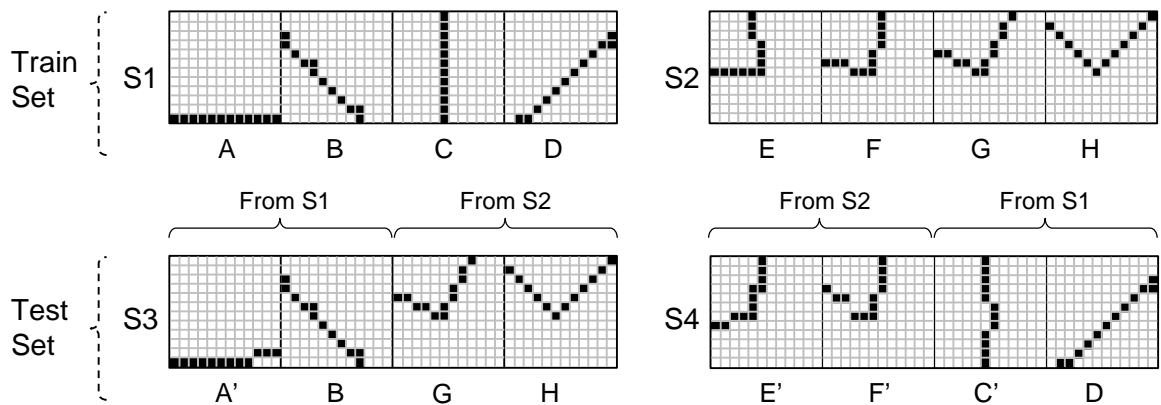

**Fig. 5** Train and test sequences used to demonstrate frame-by-frame update of the likelihood distribution for the case of spatiotemporal (sequential) patterns. The prime mark (') is used to indicate that the frame is a noisy/deformed version of the corresponding learning frame

Fig. 6 shows the state of the model, i.e., the input and the code activated in response, at the four moments of the test presentation of S1. A sample of the active afferent U (blue) and H (green) weights on each moment (frame) are shown. The cell at the source of an H weight will have been active on the prior moment. In panels, c and d, we show, for the selected cell, *all* U weights increased *throughout* learning: thus, it can be seen that the selected cell was active not just on the moment depicted [blue lines originating from active (black) pixels] but on other moments as well [blue lines originating from inactive (white) pixels]. This figure also introduces our *moment* notation. By 'moment', we mean a particular spatial input (sequence item) in the *context* of the full item sequence (prefix) that preceded it, which we indicate by enclosing the sequence including the current item is brackets and bolding the current item.

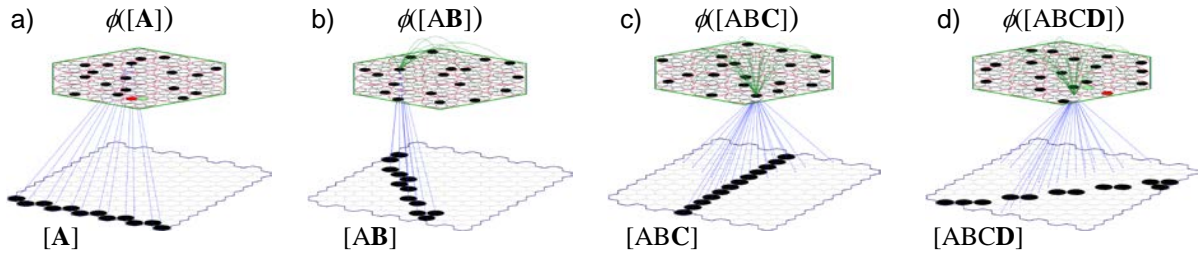

**Fig. 6** The state of the model at the four moments of the test presentation of S1.

Fig. 7 shows the processing details when the training sequence S1 is presented as a test. Note that this model has $Q$=19 CMs each with $K$=8 cells. [Note that the $U$ and $V$ traces appear separately: while this is redundant for the first item of a sequence, since $U$=$V$, it is not for all non-initial items, since $V$=$HU$]. In total, eight spatiotemporal moments are stored in the mac during learning. S1's first moment is the presentation of its first item, A, denoted [**A**], followed by [**AB**], [**ABC**], and [**ABCD**], as shown in Fig. 6. Similarly, for S2's moments, [**E**], [**EF**], [**EFG**], [**EFGH**], which are not shown

The main message of Fig. 7 is that, with each successive item presented, the likelihoods of the eight stored hypotheses, i.e., the hypothesis that the current input moment is [**A**], that it is [**AB**], etc., are updated in a way that respects the *coarsely ranked spatiotemporal* similarity structure of the experienced inputs. By "coarsely ranked", we mean the following. As this is an exact repeat of a training instance, the most likely moment at each time step is the one that occurred during the training instance. A glance down across the four likelihood charts at right verifies this: the code of the correct moment (blue bar) is activated more strongly than all others (cyan bars). In general, on each time step, the likelihoods of the other seven moments are much lower, i.e., falling into a second coarse rank. However, the distribution on the third time step (Fig. 7c) is quite plausibly described as having three ranks, the middle one including the bar for moment [**E**]. This is appropriate and is due to the fact that item E has significant overlap with C (4 pixels, see Fig. 5). When [**E**] was presented as the first item of S2's learning trial, the prior learning in the U weights that had occurred on the third moment of S1's learning trial, [**ABC**], caused high $u$ summations for the cells in $\phi$([**ABC**])*).* Since [**E**] is the first item of S2, no H signals are present and the choice of cells depends only on the U inputs, consequently yielding a relatively high intersection, $\left\{ \phi([ABC]) \cap \phi([E]) \right\}$
.

Fig. 7's charts show that the correct tracking of likelihood is achieved via *independent* soft max choices in each of the $Q$ CMs. In panels b-d, in all CMs, the correct cell has $U$=1 and $H$=1, which yields $V$=1. This reflects that fact that the test sequence here is an exact duplicate of one that has been learned, S1. Many other cells have *either* a significant $U$ *or* a significant $H$ value, reflecting *crosstalk* due to some cells being involved in the codes of multiple moments, but not both. In fact, all incorrect cells appropriately have zero or near-zero $V$ values. Thus, in all CMs, the $\rho$ distribution greatly favors the correct cell. Two errors occur, one in CM 17 for moment [**A**] and one in CM 11 for moment [ABC**D**]: as winners are picked using *soft max*, there are occasional instances in which a far less likely neuron is selected. Nevertheless, we see that the effect of global familiarity, $G$, which equals one at all four moments, modulates the $V$-to-$\mu$ transform in such a way as to cause almost the whole stored code to activate correctly, i.e., to increase the correlation of those cells. And, as will be seen in Figures 8 and 9, lower $G$ values decrease correlation. *Thus, our theory provides a novel, causal, indeed normative, explanation of correlation in the brain.* The degree of correlation from one moment to the next, both during learning and retrieval (inference), is effectively modulated (though indirectly) by a deterministic mechanism, the modulation of the $V$-to-$\mu$ transform. Indeed, all CSA steps except for the last, are deterministic.

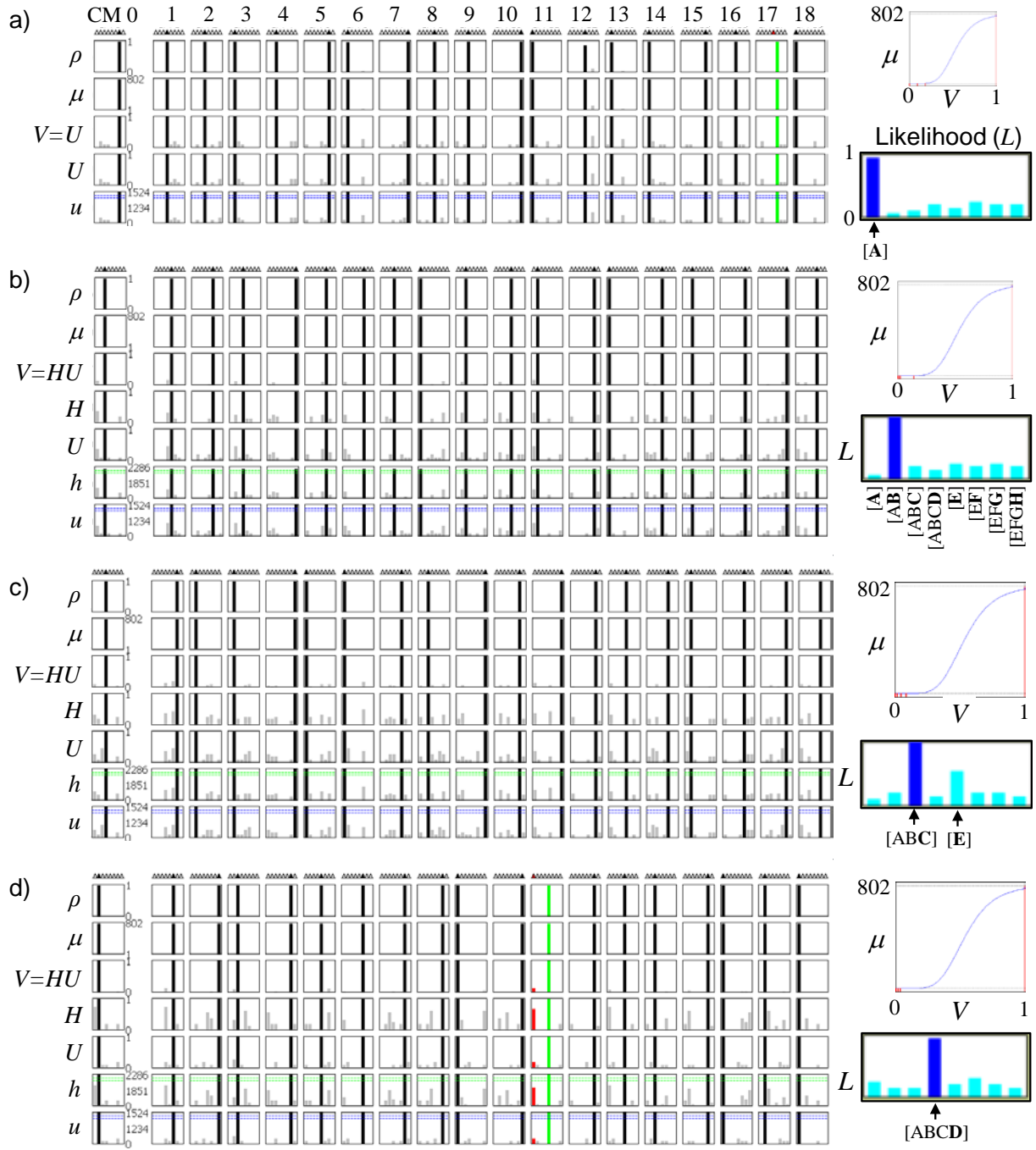

**Fig. 7** Detailed trace information for test presentation of all four items of S1=[ABCD], i.e., a "train=test" sanity test. Because S1 was also a training item and because only one other sequence has been stored in the mac (and thus, crosstalk is low), the model detects $G$=1.0 (100% familiarity) for each item and reinstates the traces almost perfectly. The $V$-to-$\mu$ function is the same at all four moments because $G$=1 in all four cases. We show the details of the $V$-to-$\mu$ transform for one of the CMs, CM 0, at upper right of each panel. The likelihoods over the eight moments stored in the mac are shown at right: the key for the eight moments is the same for all panels and is shown only in panel b to reduce clutter. Panel a has no $h$ or $H$ traces since they are not relevant on the first item of a sequence

We now consider two novel sequences as further evidence that hypothesis likelihoods, measured as active fractions of their SDR codes, track the coarsely-ranked spatiotemporal similarity of the presented sequence from moment to moment. Fig. 8 shows the state of the model at the four moments of both sequences, S3=[A'BGH] and S4=[E'F'C'D]. As noted earlier, these sequences were constructed so that the relative spatiotemporal similarities of the stored (learned) moments and the current test moment would be clear even without an exact spatiotemporal similarity metric. That is, the first two moments of S3 must clearly be considered closest to the first two moments of S1. The third moment, [A'B**G**] is *spatially* closest to the third moment of S2, [EF**G**], but from a *spatiotemporal* perspective, i.e., considering the immediately prior two moments as context, the model should reasonably consider the learned moment, [AB**C**] to also have elevated likelihood. The likelihood panel in Fig. 9c indeed shows that the two moments, [AB**C**] and [EF**G**] have the two highest likelihoods. Their precise likelihoods vary across test instances, but they are almost always the two highest. This behavior is due to flattened distributions, resulting from the lower $G$ (=0.478) and the fact that in many CMs, the cell that was in $\phi$[AB**C**]) and the cell that was in $\phi$[EF**G**]) are tied or nearly tied for the max $V$.

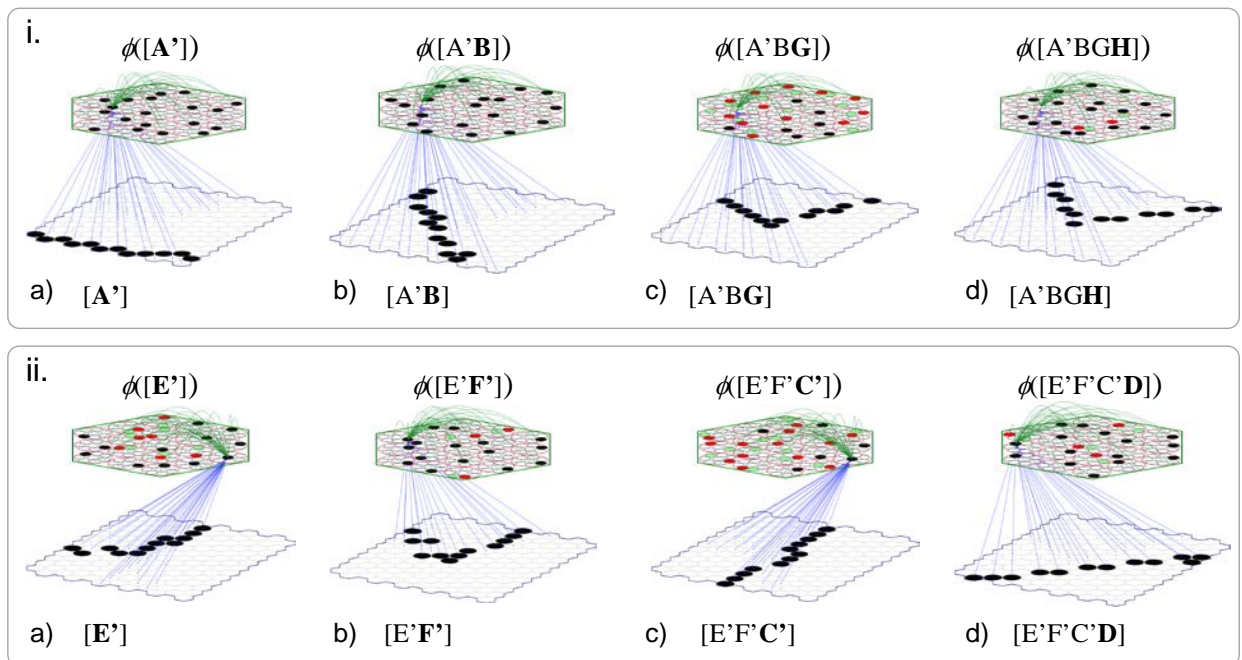

**Fig. 8** The state of the model at the four moments of each of the novel test sequences, S3=[A'BGH] (i) and S4=[E'F'C'D] (ii)

The same reasoning that allows us to consider S3's third moment, [A'B**G**] ambiguous, i.e., taking temporal context into account, suggests that the fourth moment, [A'B**GH**], should be judged less ambiguous and in fact more likely to be an instance of learned moment, [EFG**H**]. This is indeed reflected in the likelihood panel in Fig. 9d. The higher global familiarity, $G$=0.566, results in a more expansive $V$-to-$\mu$ transform, which, in combination with higher $V$ values for the cells in $\phi$[EFG**H**]) results in those cells being greatly favored in each CM. Thus, the model is seen to have successfully gone through an ambiguous state of a sequence, recovering via the on-line combining of new evidence to yield an appropriately less ambiguous internal state.

This example underscores another crucial capability of the model: namely, that by allowing stored hypotheses to be physically active (in proportion to their code's overlap with the currently active code), it allows transiently weaker hypotheses to recover based on future evidence and countermand transiently stronger hypotheses. For example, in Fig. 9c, the strongest hypothesis, [AB**C**] is consistent with the overarching hypothesis that the currently unfolding sequence is [ABCD] despite the inconsistent evidence presented on the third time step, input state G, which is more consistent with the overarching hypothesis that the unfolding sequence is [EFGH]. However, when additional evidence inconsistent with [ABCD] occurs on the fourth time step (Fig. 9d), the overarching hypothesis that the currently unfolding sequence is [EFG**H**] becomes strongest.

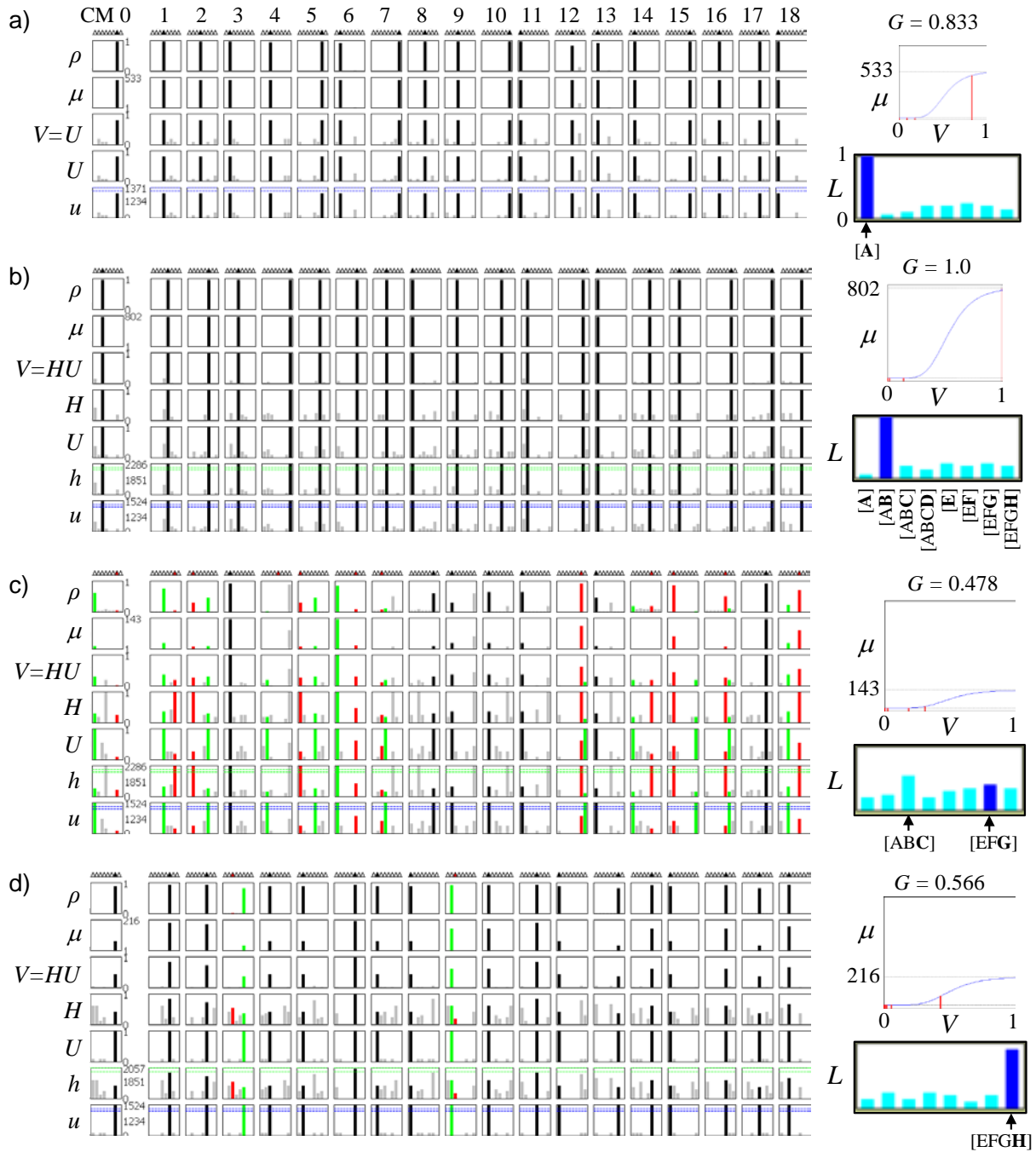

**Fig. 9** Detailed trace information for test presentation of all four items of S3=[A'BGH]

Fig. 10 shows the results for the final example, S4=[E'F'C'D] whose first two moments were constructed to be closest to the first two moments of S2 and whose second two moments are very close to the last two moments of S1. The behavior of the model is broadly analogous to its behavior for S3 in that at its third moment, [E'F'**C'**], the two learned moments, [AB**C**] and [EF**G**], deemed most likely by the model are, by the same reasoning applied for S3, plausibly spatiotemporally most similar to [E'F'**C'**].

Global familiarity, $G$=0.314, is lower here than for the third moment of S3 because we introduced more noise to the first two moments of S4 than to S3. In fact, the spatial input E' has the same intersection with E as it does with C (as can be seen by close inspection of Fig. 5) and F' has 9 of 12 pixels in common with F. Thus, learned moments, [AB**C**] and [**E**], have approximately the same likelihood in Fig. 10a. Although the cells comprising $\phi$([AB**C**]) were originally chosen in a spatiotemporal context and therefore had their afferent H weights increased [from the cells comprising $\phi$([A**B**])], there are no H signals present on the first item of any sequence. Thus, the choice of winners on the first moment of S4 depends only on U signals and thus, on the increases made to the U weights during learning. This is why learned moment [AB**C**] receives a high likelihood at this moment.

Fig. 10b shows that the ambiguity present on the first moment is greatly diminished by the presentation of spatial input F', which is quite similar to F, thus making the spatiotemporal input moment, [E'**F'**], much more spatiotemporally similar to stored moment, [E**F**], than to stored moment, [ABC**D**], or to any other learned moment. On the third moment of S4 (Fig. 10c), we present spatial input C' which is much more similar to C than any other spatial input. As was the case for the third moment of S3, this leads to the two learned third moments, [AB**C**] and [EFG] being approximately equally likely. However, due to the increased noise present in the first three moments of S4 compared to S3, $G$ is lower here and the likelihoods of these two learned moments, i.e., the fractions of their codes active, are appropriately lower than was the case on the third moment of S3. Finally, we present spatial input D on the fourth moment, [E'F'C'**D**]. Due to the multiplication of the U signals from D and the H signals from $\phi$([F'**G'C'**]), which had appreciable overlap with $\phi$([AB**C**])], the cells comprising $\phi$([ABC**D**]) are highly favored in all CMs and end up winning in most of them, yielding the significantly higher likelihood for [ABC**D**] than all other moments, as seen in Fig. 10d.

Again, the model is seen to negotiate ambiguous moments, updating its active code from moment to moment, such that it simultaneously represents what are plausibly (given its small experience of the world) the most likely hypothesis (or hypotheses) and the full coarsely-ranked distribution over hypotheses. It has been pointed out that this ability, i.e., simultaneously representing multiple competing hypotheses, as for example is required for representing motion transparency, is problematic for theories which use fully distributed codes, as do the PPC theories (Pouget, Dayan et al. 2000). It is true that we do not go into numerical detail justifying most of the relative similarities/likelihoods (pair-wise and higher orders) present in the likelihood distributions of Figures 7, 9, and 10, nor, for that matter, those present in the spatial examples of Figures 3 and 4. However, our examples do demonstrate coarse correlation of spatial / spatiotemporal similarity with likelihood (measured as active fraction of code). Similar capabilities, such as being able to store and successfully recognize/retrieve large numbers of complex sequences, in which the same item(s) may occur multiple times and in varying contexts (e.g., a natural lexicon), have previously been demonstrated (Rinkus 1996, Rinkus 2014).

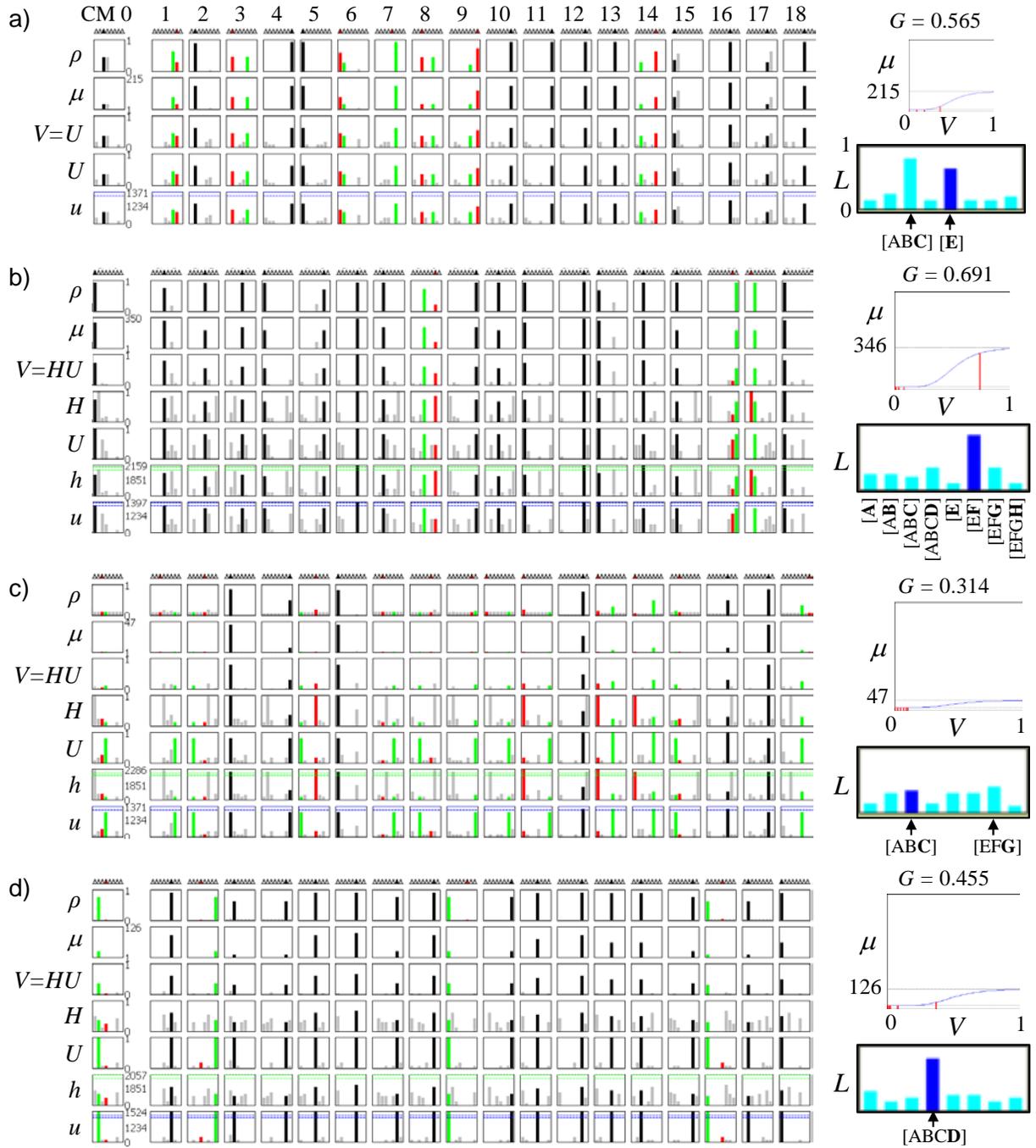

**Fig. 10** Detailed trace information for test presentation of all four items of S4=[E'F'C'D]

## 3    Discussion

We described a radically different theory, from the prevailing probabilistic population coding (PPC) theories, for how the brain represents and computes with probabilities. This theory avails itself only in the context of *sparse distributed representation* (SDR), as opposed to the *fully* distributed coding context in which the PPC models have been developed. The theory, Sparsey, was introduced 20+ years ago [originally called TEMECOR (Temporal Episodic Memory using Combinatorial Representations)], as a model of the canonical cortical circuit and a computationally efficient explanation of episodic and semantic memory for sequences, but its interpretation as a way of representing and computing with probabilities was not emphasized. The PPC models (Georgopoulos, Kalaska et al. 1982, Zemel, Dayan et al. 1998, Pouget, Dayan et al. 2003, Sanger 2003, Jazayeri and Movshon 2006, Ma, Beck et al. 2006, Rajkumar and Pitkow 2016) share several fundamental properties: 1) continuous (graded) neurons; 2) *all* neurons formally participate in every code; 3) due to 1 and 2, synapses must either be graded or rate-coding must be used to allow decoding; 4) generally assume rate-coded signaling; 5) individual neurons are generally assumed to have unimodal, e.g., bell-shaped, tuning functions (TFs); 6) individual neurons are assumed to be noisy, e.g., firing with Poisson variability; and 7) this noise and correlation, i.e., noise correlation, is generally viewed as degrading computation and needs to be mitigated, e.g., averaged out.

In contrast to these PPC properties/assumptions, Sparsey assumes: 1) binary neurons; 2) individual codes are small (relative to whole coding field) sets of cells (SDRs) and any such code *simultaneously* represents both the best matching stored hypothesis and the similarity (and thus likelihood / probability) distribution over *all* stored hypotheses; 3) only effectively binary synapses; 4) signaling via waves of contemporaneously arriving first-spikes from afferent SDR codes; 5) all afferent weights to coding neurons are initially zero, i.e., the TFs are initially completely flat, and emerge via single/few-trial learning to reflect a cell's specific history of inclusion in codes; 6) neurons are not assumed to be intrinsically noisy and are simply on or off on any given computational cycle (we do not require a spiking model); and 7) noise is a resource that is explicitly generated and injected into the code selection process to achieve a specific coding goal, namely that the overall set of codes stored in a coding field has the SISC property, which manifests, indirectly, in particular patterns of correlation amongst the individual units. Thus, Sparsey entails a completely different view of noise/correlation from the mainstream. Rather than being viewed as a problem imposed by externalities (e.g., common input, intrinsically noisy cell firing), it essentially functions in a positive sense, i.e., as a resource.

Specifically, we showed that: i) if a model uses SDR coding; ii) if the model's process of assigning SDR codes preserves similarity from the input space into the code space (the SISC property); and iii) if input similarity can be assumed to correlate with likelihood, then:

a) The active SDR code simultaneously represents *both* the most probable hypothesis and the likelihood/probability distribution over *all* stored hypotheses. Specifically, the likelihood/probability of any hypothesis is represented by the *fraction* of its code's cells that is active, as part of the current, fully active code.

b) In the spatiotemporal (sequential) case, with each successive sequence item, Sparsey's core algorithm, the *Code Selection Algorithm* (CSA) (Table 1), updates the entire distribution [cf. belief update (Pearl 1988)] in approximate accord with intuitive notions of spatiotemporal similarity and with a number of computational steps that remains fixed as the number of stored hypotheses (SDR codes) increases.

That is, executing the CSA is dominated by a single iteration over the weights (Step 1) the number of which is fixed for the life of the system. We emphasize that this algorithmic efficiency for both learning and retrieval has not been shown for any other computational method, including hashing methods, either neurally-relevant (Salakhutdinov and Hinton 2007, Salakhutdinov and Hinton 2009, Grauman and Fergus 2013), or more generally [reviewed in (Wang, Liu et al. 2016)]. Although time complexity considerations like these have generally not been discussed in the PPC literature, they are essential for evaluating the overall plausibility of models of biological cognition, for while it is uncontentious that the brain computes probabilistically, we also need to explain the extreme speed with which these computations, which are over potentially quite large hypothesis spaces, occur.

The key to Sparsey's computational speed is its extremely efficient method of computing the *global* familiarity, $G$, of an input (to a mac), and using $G$ to adjust an individual cell's transfer function from its own local similarity measure, $V$, to its final probability, $\rho$, of being chosen winner [in its own winner-take-all competitive module (CM)]. $G$ can be viewed as directly modulating the noise present in the process of code selection. That is, when high familiarity is detected, noise is minimized by disproportionately increasing the bias towards selecting cells that are more correlated with input (higher correlation being indicated by higher $V$), i.e., *pattern completion*, and when low familiarity is detected, noise is maximized by making win probability of all cells in each CM more equal (when $G=0$, all cells are equiprobable), i.e., *pattern separation*. See (Rinkus 2010) for a sketch of a possible noise modulation mechanisms involving one or more of the brain's neuromodulators. We emphasize that this mechanism constitutes a radically novel concept of noise and correlation in the brain. Moreover, it constitutes a novel method for combining global and local information during inference (and learning). In particular, it suggests the possible necessity of a structural *mesoscale* (in our case, the WTA CM) to facilitate the use (mixing) of global information into the local decision processes. From the opposite standpoint, $G$'s action can also be viewed as controlling the amount of correlation amongst the neurons, or relatedly, as controlling which cells are *bound* together to represent inputs, either purely spatial inputs or spatiotemporal (sequential) events, and thus providing similar functionality to binding operations described in (Kanerva 1994, Plate 1997, Rachkovskij and Kussul 2001, Kanerva 2009).

In most other SDR models, the coding field is a homogenous field of binary units from which some number are chosen to be in any particular code (Kanerva 1988, Moll and Miikkulainen 1997, Rachkovskij and Kussul 2001, Hecht-Nielsen 2007, Snaider and Franklin 2011, Snaider and Franklin 2012, Snaider and Franklin 2012, Snaider and Franklin 2012). This is also true of *combinatorial neural codes* (Willshaw, Buneman et al. 1969, Osborne, Palmer et al. 2008, Curto, Itskov et al. 2013) as well as for all the binary hashing models reviewed in (Wang, Liu et al. 2016). In contrast, in Sparsey, the coding field consists of $Q$ winner-take-all (WTA) competitive modules (CMs), each comprised of $K$ binary units. Selecting a code is performed by making $Q$ independent draws, one in each of the $Q$ CMs. Thus, unlike these other homogenous-field models, Sparsey has an *explicit structural mesoscale*, the WTA CM, situated between single neuron and the whole coding field. In fact, from its inception, Sparsey has been offered as a generic model of the cortical macrocolumn, with its WTA CMs proposed as analogous to minicolumns (Rinkus 1996, Rinkus 2010, Rinkus 2014). Hence, our synonymous use of "coding field" and "macrocolumn", or just, "mac". One important consequence of the presence of this explicit *structural* mesoscale is that it imposes a specific, fixed sparsity *structurally*. Thus, no explicit computation (or the energy consumed by it) need be expended to control sparsity during the model's lifetime, during either learning or retrieval (inference). This contrasts with the far more prevalent technique of adding a penalty term to a cost function to achieve sparsity in "sparse coding" models (Olshausen and Field 1996, Perrinet 2015), which does entail continual computation throughout model lifetime.

We point out that one other SDR-based model, Numenta's *hierarchical temporal memory* (HTM) model (Ahmad and hawkins 2015, Cui, Ahmad et al. 2016), does have a mesoscale, which is also equated with the cortical minicolumn, however HTM's conception of the minicolumn differs radically from ours. In particular, in HTM, all cells in a minicolumn have the same TF *a priori* and all cells in a minicolumn co-activate when the appropriate feature is present in the minicolumn's RF. While it is true that Hubel and Wiesel's original results found that all cells along vertical penetrations through visual cortex had similar TFs, more recent studies with more detailed probes and more refined observation methods are revealing more heterogeneous TFs than originally thought. This suggests that our approach in which TFs are *learned* from scratch and can end up being arbitrarily heterogeneous may be have wider applicability. In addition, HTM's assumption that *all* cells in a minicolumn, or even in, say, the L2/3 volume of a minicolumn, activate simultaneously seems clearly at odds with experimental data such as calcium imaging, e.g., (Ohki, Chung et al. 2005). Moreover, although HTM uses fixed-density SDRs, they are not imposed structurally and thus, in contrast to Sparsey, *do* require explicit computation (and energy use) to determine which subsets of minicolumns will activate in any given instance, presumably during both learning and retrieval. One additional, important point of distinction is that while HTM, like Sparsey, does possess a mesoscale, to our knowledge, all published HTM results thus far involve only single SDR coding fields (which they term "regions"), whereas (Rinkus 2014) describes results of hierarchical Sparsey models consisting of multiple internal levels, each consisting of multiple coding fields (macs). While the mesoscale (minicolumn) architecture is functionally crucial, there is substantial evidence for the subsuming macrocolumn-scale in various cortical regions/species and there are crucial functional advantages associated with that scale as well.

In addition to Sparsey's efficiency from a purely algorithmic standpoint, we emphasize that because it requires only binary neurons and synapses, it does not require rate coding. Rather, it is naturally suited to signaling via waves of contemporaneously arriving first-spikes from one SDR code to the next, either recurrently or to downstream coding fields. Thus, rather than the ~100 ms that reliably decoding spike frequency requires, our model requires only a few ms window during which contemporaneous signals from an afferent SDR code might arrive at a downstream coding field and be integrated. We imagine that some sort of macrocircuit-level control apparatus, e.g., integration during some phase of a gamma-scale envelope [cf. (Buzsáki 2010)], might impose such a window, a hypothesis we would like to explore in the future. We also emphasize the potentially significantly lower metabolic/energy costs of signaling based on first-spikes compared to signaling via rate-coding (in which many spikes must be sent and integrated). This is in addition to the already reduced energy costs of sparse coding compared to dense coding.

Sparsey is an existence proof that representing and using graded values, e.g., probabilities, requires only binary neurons and binary synapses. There is no need for explicit *localist* representation of graded values, anywhere in the system/computation. More specifically, there is no need to represent probabilities/likelihoods as spatially localized firing rates (i.e., at particular synapses) and there is no need to represent conditional probabilities, or strengths of association, via continuous/graded weights. Moreover, we contend that replacing what, in a localist model, is typically a *single* real valued parameter representing the relation, e.g., conditional probability, between two symbolic-level variables, with a *bundle*, i.e., "synapsemble", of *independent* binary parameters, allows more flexible and faster learning of an input domain's statistics. Developing this argument is one of our near-term future research goals.

As Sparsey is a distributed memory, the traces are stored in superposition and therefore interference (crosstalk) does increase with the number of hypotheses stored. For a given setting of parameters, there will be regime in which the number of stored hypotheses is low enough so that expected interference, i.e., expected retrieval error, remains tolerable. If the overall Sparsey system had only one internal level and only one mac at that level, then analyses characterizing capacity, expected errors, vis-à-vis parameters, would be the primary research focus. But this is not the case. The single Sparsey module, the mac, is proposed as the analog of the cortical macrocolumn. And, the cortex is known to be organized as a deep hierarchy, with perhaps >10 cortical levels (along some paths), where each level is a patch of order hundreds to thousands of macs. The question of overall storage capacity as it relates to explaining the apparent vast storage capacity of the typical human over most of his lifespan will thus depend on how information is distributed throughout and dynamically interacts across the entire hierarchy, not just in a single mac.

The advantages of organizing knowledge hierarchically, both categorically and componentially (part-whole) have long been known. More recently, the advantages of many-leveled vs. flat representations have been described in terms of the efficiency (essentially, the number of parameters needed) of representing highly nonlinear relations (Bengio 2007, Bengio, Courville et al. 2012) and the constant stream of impressive "Deep Learning" results strongly bears this out (Krizhevsky and Hinton 2011, LeCun, Bengio et al. 2015, Silver, Huang et al. 2016). However, Deep Learning models, including LSTM (Hochreiter and Schmidhuber 1997), have thus far not been combined with SDR, and indeed, the principles of the two paradigms are very different and may be essentially incompatible. In this regard, we must specifically point out that the recently described Sparsely-Gated Mixtures of Experts (MoE) model (Shazeer, Mirhoseini et al. 2017), while exploiting a principle of sparsity (referred to more generally as *conditional computation* in which the explicit goal is to minimize the fraction of the machines parameters participating in any given computation), is not an instance of SDR as present in Sparsey or other SDR models. One important difference with Sparsey is that the Sparsely-gated MoE does not use the content (statistics/semantics) of the input to select which experts respond to and are thus used to code the input. More generally, this is true of many of the instantiations of the "drop-out" principle (Bengio 2013, Srivastava, Hinton et al. 2014). In contrast, as we have explained throughout, Sparsey's CSA implements a spatial/spatiotemporal (and in principle, multimodal) matching mechanism that directly uses the input to control which cells code for an input, which yields the crucial SISC property.

We have also been exploring multi-level hierarchies, but of SDR coding fields, e.g., the model used in our explanation of single-cell TFs, and will be continuing on this path. We believe that truly capturing the essence of how the brain computes requires the union of hierarchy/heterarchical organization and *sparse* distributed coding, which places our work in distinction with, on the one hand, Deep Learning models, which combine *densely/fully* distributed coding fields (i.e., Boltzmann or MLP fields) with hierarchy [though (Shazeer, Mirhoseini et al. 2017) appears to be an exception], and on the other hand, HMAX models (Riesenhuber and Poggio 1999, Riesenhuber and Poggio 2002, Serre, kouh et al. 2005, Poggio and Serre 2013), which combine *localist* coding fields with hierarchy.

We point out that there is another, new theory of probabilistic computation in the brain, for which distributed representation is essential, and which is much closer in spirit to Sparsey (Rajkumar and Pitkow 2016). Despite the fact that their theory shares most of the PPC properties mentioned at the outset, at a high level, Sparsey can be described in terms quite similar to those used by (Pitkow and angelaki 2016), which describes their theory as having three main parts: a) overlapping patterns of population activity are proposed to encode the latent variables of the observed domain; b) the brain specifies how those variables are related to the world through a sparse probabilistic graphical model; and c) recurrent circuitry implements a nonlinear message-passing algorithm that effectively executes probabilistic inference amongst the latent variables represented in the population codes residing in a hierarchy of coding fields. The analog of point (a) in a Sparsey mac is that the 'overlapping patterns of population activity' correspond to subsets of cells, i.e., intersections, that occur amongst multiple codes, i.e., across multiple contexts. These subsets are of size smaller than the whole code size, $Q$, and crucially, they emerge over the course of learning. Furthermore, there will in general be intersections over these subsets as well, allowing for the encoding of statistics of a range of higher orders. The analog of point (b) is that the modifications made to Sparsey's synaptic projections, including the recurrent projection of a mac to itself and its U, H, and D, projections to/from other macs at its own and other levels of the hierarchy embed the probabilistic relations amongst the latent variables. Sparsey's analog of point (c) is essentially the operation of the CSA. These characteristics mean that these models can be viewed as distributed instantiations of graphical probability models (GPMs), which by large, have been localist, e.g., hidden Markov models, Bayesian nets, dynamic Bayesian nets. We believe that the move from a localist GPM concept to a distributed GPM concept has major implications, notably: i) that the domain's latent variables (their identities and valuednessess) are mapped to distributed codes which emerge over the course of on-line learning, cf. "anonymous latent variables", (Bengio 2013); and ii) that the conditional probabilistic relations amongst the variables are represented partially in the code intersections and partially in intersections of synaptic mappings (i.e., synapsemble intersections), which also emerge over the course of learning.

After many decades, experimental methods are finally reaching the point where the fast time scale activities of all neurons in large, e.g., macrocolumn-scale, volumes will be observable. This will finally allow us to understand the brain's operation in what we believe is its native language, essentially cell assemblies and sequences of cell assemblies (Hebb 1949), where in our view, each such assembly functions both as the most likely hypothesis (stored memory) and the distribution over all memories stored (in the cell assembly's container, which we propose is the cortical macrocolumn). We offer Sparsey as a theoretical elaboration of this concept, one which is simple, i.e., binary cells and synapses, single-trial Hebbian learning (with some additions, e.g., decay, in the fully general model), and general, in that the CSA is extremely generic, and powerful, in terms of its computational efficiency. There are numerous questions to pursue, notably regarding the nature/capacities of hierarchical interactions amongst macs and through time and we look forward to continuing to explore them.

# 4    Methods

## 4.1    Similarity Metric

The similarity metric for the case of spatial inputs is simply pixel-wise overlap. If all inputs have exactly the same number of active pixels, we can measure spatial similarity simply as size of intersection divided by that number, which for the examples of Figures 3-4, is 12:

$$sim(I_x, I_y) = \left| I_x \cap I_y \right| / 12$$

For the spatiotemporal examples of Figures 5-10, we use pixel overlap as the spatial measure for each frame, individually, but use only semi-quantitative estimation regarding the temporal aspects of the sequences, as described in the main text.

The input patterns for the spatial and spatiotemporal update examples (Figs 3-10) were manually created to minimize pixel-wise overlap, but are otherwise similar to natural inputs that are subject to preprocessing consisting of edge filtering, binarization, and skeletonization. The inputs used in receptive field example were created from the KTH data set using the aforementioned preprocessing.

## 4.2    The model

The model's architecture is described in detail in the main body, as is its algorithm, the code selection algorithm (CSA) including parameter values (Table 1) and so will not repeated here.

The learning law is Hebbian: simultaneous pre- and post-synaptic activation causes the synapse to be set to its max weight, 127, which is binary "1". In the case of H and D weights, the weight is increased if the pre- and post-synaptic cells are active on successive time steps. In the full Sparsey model, additional learning principles are modeled including decay, permanence, and critical period, see (Rinkus 2014) for details. The simulations described in this paper are of small enough size, in terms of total numbers of inputs, which in the sequential case, means total number of frames (items) across all training sequences, so that these additional learning principles do not materially affect the results/conclusions.

Nevertheless, we briefly describe these additional principles. Following a synaptic weight increase (which is always to the max weight) due to a pre-post coincidence, there is an initial period in which the weight remains at or near its max, and then decays with an approximately inverse logarithmic profile. If a second pre-post coincidence occurs within a relatively small temporal window of the previous pre-post coincidence, then the weight is reset to the max value (127) and its permanence is increases, i.e., the time scale of its decay is greatly protracted. In our simulations to date, these principles have been quantified in explicit tabular forms and rule-based use of the tables, as described in (Rinkus 2014). The motivation is that the expected time of recurrence of pre-post coincidences that is due to structural regularities of the input domain must clearly be much shorter than for pre-post coincidences due to noise or spurious alignments. Thus, the described mechanism will preferentially embed SDR codes of events that are due to structural (statistical) regularities of the domain, while allowing spurious events to fade from memory. This permanence mechanism/protocol is a form of metaplasticity, similar in spirit to the Cascade model (Fusi, Drew et al. 2005), as well as to more recent attempts to deal with catastrophic forgetting (Aljundi, Babiloni et al. 2017, Kirkpatrick, Pascanu et al. 2017), but far simpler than these other models in that it does not require any explicit evaluation of a synapse's importance/relevance to a global objective, e.g., accuracy on learned tasks.

This works in concert with a critical period mechanism. As we discussed, an SDR coding field has a finite storage capacity. The code space is exponential, e.g., in Sparsey's case, $K^Q$, but as more and more codes are embedded in superposition, interference (crosstalk) accrues. As the fraction of the increased weights grows, expected retrieval accuracy falls and will go to zero if learning is not frozen, i.e., if a critical period is not enforced. In our full model, all three principles/mechanisms work in concert, but again, this paper's simulations are small enough so that none of them come into play.

### 4.3    Code Similarity (Likelihood) Metric

Given that SDR codes consist of one binary unit chosen from each of the $Q$ competitive modules (CMs) comprising the coding field, code similarity is measured as Hamming distance normalized by code size, $Q$. Given: a) our assumption that input similarity correlates with likelihood; and b) that the model preserves input similarity into code similarity, we measure the likelihood (for either the spatial or spatiotemporal, i.e., "moment" case) of the stored (learned) input $I_x$ given the current input $I_C$ as:

$$L(I_x) = \left| \phi(I_C) \cap \phi(I_x) \right| / Q$$